# THE NCS-MODEL: A SEISMIC FOUNDATION MODEL TRAINED ON THE NORWEGIAN REPOSITORY OF PUBLIC DATA

Alba Ordoñez[1], Theodor Johannes Line Forgaard[1], David Wade[2], Aina Juell Bugge[3], Håkon Nese[3], Anders Ueland Waldeland[1]

[1] Norwegian Computing Center
[2] Equinor (Norway)
[3] Aker BP (Norway)

ABSTRACT

We present the NCS-models, a family of seismic foundation models pretrained on a large curated share of full-stack seismic cubes from the Norwegian Continental Shelf (NCS) available through the public DISKOS database. The model weights are open-sourced for the wider geoscience community.

Foundation models trained with large-scale self-supervision are emerging as a promising basis for automatic seismic interpretation. However, most existing seismic models rely on limited or proprietary datasets, and it remains unclear how well natural-image foundation models transfer to seismic data. Our goals are to develop basin-scale seismic foundation models, provide practical recipes for scalable 3D training, and compare the downstream performance of practical configurations trained under similar pretraining budgets.

Using masked autoencoders with Vision Transformer backbones, we pretrain models on a DISKOS-derived corpus of 3D time- and depth-migrated seismic volumes. The NCS-model variants use 2D, 2.5D multi-view, and 3D tokenization and each variant is pretrained using approximately 250 GPU-hours on the same hardware. We evaluate the models using frozen backbones together with k-nearest neighbors and linear probing on NCS interpretation benchmarks and one out-of-basin benchmark. Baselines include an ImageNet-pretrained MAE, a frontier vision foundation model, and a globally pretrained seismic foundation model.

Following large-scale pretraining on the curated DISKOS corpus, NCS-2.5D achieves the highest average performance among all evaluated models. The resulting embeddings additionally support similarity search for interactive interpretation.

## INTRODUCTION

The widespread digitization of seismic acquisition and processing has produced vast databases of seismic images from diverse basins worldwide, creating extensive opportunities for automated analysis and data-driven subsurface interpretation. The complexity and volume of modern seismic data demand powerful computational resources and advanced algorithms to enable effective and reliable analysis. In recent years, deep learning (DL) has emerged as a powerful framework capable of addressing these demands, enabled by advances in modern high-performance computing offering scalable, data-driven approaches for extracting meaningful structure and patterns from large seismic datasets. DL-based methods now play an increasingly central role in seismic interpretation, supporting tasks such as fault detection, horizon tracking, facies classification, and general seismic pattern recognition (Waldeland et al., 2018; Wu et al., 2018; Zhao, 2018; Dramsch, 2020; Wrona et al., 2021; Yang and Sun, 2020; Fernandes et al., 2024). These developments have demonstrated the potential of DL to enhance structural and stratigraphic understanding while reducing manual workload and enabling more consistent, scalable interpretation workflows. However, classical DL algorithms require large amounts of labelled data for training and may not generalize well between surveys.

A major development in the DL landscape is the emergence of foundation models (FMs): large, pretrained neural networks that learn general-purpose representations from unlabeled data, enabling transfer with significantly less training to a wide range of different use cases, commonly referred to as downstream tasks in the computer vision community (Radford et al., 2021; Bommasani et al., 2022). FM frameworks such as MAE (He et al., 2022), the DINO family models (Caron et al., 2021; Oquab et al., 2023; Siméoni et al., 2025), and multimodal systems like CLIP (Radford et al., 2021) have achieved state-of-the-art transfer performance when trained at web scale (e.g., ImageNet, (Deng et al., 2009); Instagram-3.5B, (Mahajan et al., 2018); or billions of image-text pairs (Schuhmann et al., 2022)). These frameworks typically employ the Vision

Transformer (ViT) (Dosovitskiy et al., 2021) as the underlying image encoder due to its strong scaling behavior and suitability for self-supervision.

However, applying these models directly to seismic data exposes fundamental mismatches between the natural image domain and the seismic data domain. First, we are imaging subsurface structures which are inherently different from objects patterns typical of natural images. In addition, the generative processes behind seismic amplitudes differ fundamentally from natural-image formation: they arise from elastic wave propagation and subsurface reflectivity ((Yilmaz, 2001; Aki and Richards, 2002)), are shaped by acquisition geometry (Claerbout, 1985), and are further modified by processing pipelines. As a result, the invariances and spatial statistics encoded in natural-image FMs may not necessarily transfer to seismic interpretation tasks.

These differences motivate the development of seismic-specific pretraining strategies, where models learn representations directly from collections of seismic cubes and are tailored to the structural and stratigraphic patterns relevant for interpretation. This perspective has been explicitly articulated in recent geophysical work (Liu et al., 2025).

In the seismic context, the emergence of foundation models has been consolidated in a review by (Fuchs et al., 2025), which surveys foundation-model approaches for seismic data processing and discusses how choices in pretraining strategy and architecture shape performance and practicality. Notably, the review identifies masked autoencoding, following the MAE framework of (He et al., 2022) implemented with ViT backbones, as a prevalent strategy that has contributed to articulating the concept of seismic foundation models. Within this line of work, (Sheng et al., 2025) formalize the notion of seismic foundation model training and demonstrate transfer to multiple downstream tasks using representations learned through MAE pretraining. They also provide an open-source release of models, training workflows, and data. Using a similar pretraining strategy and architecture, (Pham et al., 2025) introduce SeisBERT, further demonstrating the transferability of learned representations across tasks and improvements over DL-based baselines. Finally, (Sansal et al., 2025) investigate the role of scale, providing empirical evidence that increasing data volume and model capacity improves performance, and highlighting the importance of infrastructure and data handling when operating in 3D, in contrast to (Sheng et al., 2025) and (Pham et al., 2025), both of which focus on 2D data.

These studies establish the promise of seismic foundation models, while leaving open questions. First, some of the models and datasets are not publicly accessible, preventing independent validation or reuse. Second, when models are released, they are often trained on small datasets (<2GB), making it unclear whether they capture the true geological variability encountered in practice.

In addition, existing seismic FMs are typically trained or adapted to globally aggregated datasets (Gao et al., 2025; Dou et al., 2025; Sheng et al., 2025). While such corpora can be valuable for learning generic seismic primitives, they rarely contain sufficient coverage of any single basin to represent its geologically coherent variability at scale. On the Norwegian Continental Shelf (NCS), for example, interpreters encounter a mixture of rift-related faults, salt diapirs, injectites, glacial erosion surfaces, and complex stratigraphic geometries. In principle, an ideal global FM would include substantial NCS representation and thereby capture these patterns. However, NCS data are usually sparse within global training corpora, which could limit the model's ability to learn basin specific structure and stratigraphy. This motivates training on this archive to examine the usefulness of representations learned from extensive regional coverage and evaluate whether we could emphasize patterns relevant to local structures.

We therefore ask how foundation models pretrained on the NCS archive perform relative to available natural-image and seismic baselines.

In this paper, we introduce the NCS models, a new family of seismic foundation models trained exclusively on real-world 3D seismic cubes from the Norwegian Continental Shelf. The training corpus comprises 829 full-stack seismic cubes totaling 27 TB, spanning multiple acquisition survey vintages, acquisition styles, processing workflows, and data quality present on the shelf. Using the ViT-MAE framework of He et al. (2022), we investigate standard 2D tokenization, the 2.5D multi-view formulation of Waldeland et al. (2025), building on Ordoñez et al. (2024), and 3D volumetric tokenization. The models share a ViT-Base backbone design and are trained under similar pretraining budgets, while differing in input representation, positional encoding, and some training settings.

The purpose of this study is threefold:

(1) to develop regional seismic foundation models using a large archive of real industrial data from the NCS;

(2) to compare feasible 2D, 2.5D, and 3D configurations using comparable training resources and frozen-feature evaluation;

(3) to release the trained models and point to benchmark datasets to promote transparency, reproducibility, and further research in seismic FM development.

The remainder of this paper is structured as follows. Section 2 describes the NCS dataset and its geological and acquisition diversity. Section 3 presents the model architectures and training strategy. Section 4 outlines our evaluation methodology. Sections 5, 6 and 7 report benchmark results, model characteristics and practical interpretation examples based on interactive similarity search. Finally, Section 8 discusses the findings, states the limitations, and outlines directions for future work.

## THE NCS TRAINING DATASET

The Norwegian Continental Shelf is one of the most comprehensively imaged offshore regions in the world, with more than five decades of continuous seismic acquisition supporting exploration, production and monitoring activities. Central to this data ecosystem is DISKOS, Norway's national petroleum data repository, which provides standardized access to seismic, well, and production datasets released by the Norwegian Petroleum Directorate. For all conducted surveys, the seismic cubes are uploaded to DISKOS and are, after a certain confidentiality period, made publicly accessible to DISKOS member companies. DISKOS currently hosts over 1,300 public 3D seismic surveys acquired by multiple operators since the early 1980s, making the NCS uniquely suited for large-scale data-driven research and the development of region-specific FMs. In addition to pre-stack data, different stacks (full-stack, various angle stacks) with different processing and migration methods of each survey are uploaded, as well as some attribute cubes. As of June 13, 2025, DISKOS hosts more than 24,000 public post-stack cubes.

Selecting training data is a balance between selecting an amount that is practical to process while maintaining a high variability in the training set. Blindly selecting cubes could even reduce the variability. To train the NCS models, we focused on full-stack cubes and included both time-migrated and depth-migrated data. From the public repository, we aimed to select one full-stack depth cube and one full-stack time cube per survey. Stack type, migration method, and domain are recorded in the cube database metadata, but only as unstructured free-text in a description field rather than in a standardized, structured format. We therefore relied on automatic filtering of the metadata to determine stack type and domain. We made sure to cover a wide range of processing and migration methods to ensure the models applicability across different processing methods, but we did not want to select more than one or two cubes per survey to reduce redundancy in the distribution of geology. We excluded vintages from 4D surveys, to avoid having nearly fully redundant cubes in our training set. Finally, we manually inspected a central inline and a central crossline from each cube to exclude the ones dominated by noise and to remove any non-seismic cubes that were not eliminated by earlier filtering. The resulting dataset comprises 829 seismic cubes, totaling 27 TB.

Figure 1 summarizes the spatial distribution of the selected 3D seismic cubes. Survey coverage is highly heterogeneous across the NCS, with the greatest overlap concentrated along the North Sea and Norwegian Sea margin (Figure 1a), reflecting decades of intensive exploration and field development. Coverage becomes progressively sparser both toward the Barents Sea (Figure 1b) and as one move farther offshore away from the coast. This basin-scale heterogeneity, in both spatial coverage and survey redundancy, offers a rich training substrate but also underscores the challenges of working with national repository scale data, where models can easily be overfit to data-rich areas. Figure 2 shows the size of the selected surveys as a function of acquisition year. Although the repository is most densely sampled in the 2005 to 2020 interval, cube sizes still span roughly three orders of magnitude across all vintages, reflecting the coexistence of small targeted surveys and large regional surveys and underscoring the strong heterogeneity of the DISKOS repository.

To illustrate the range of imaging quality across acquisition vintages and subsurface settings, Figure 3 shows three representative inline sections selected from different surveys. These examples range from severely degraded data affected by acquisition footprint, multiples, or missing frequency content (Figure 3a), to legacy surveys with moderate noise and limited bandwidth (Figure 3b), and high-fidelity broadband acquisitions with clear reflector continuity (Figure 3c). This variability is characteristic of the DISKOS repository and underscores the importance of training foundation models on data that span the full spectrum of real-world conditions encountered in industrial interpretation workflows.

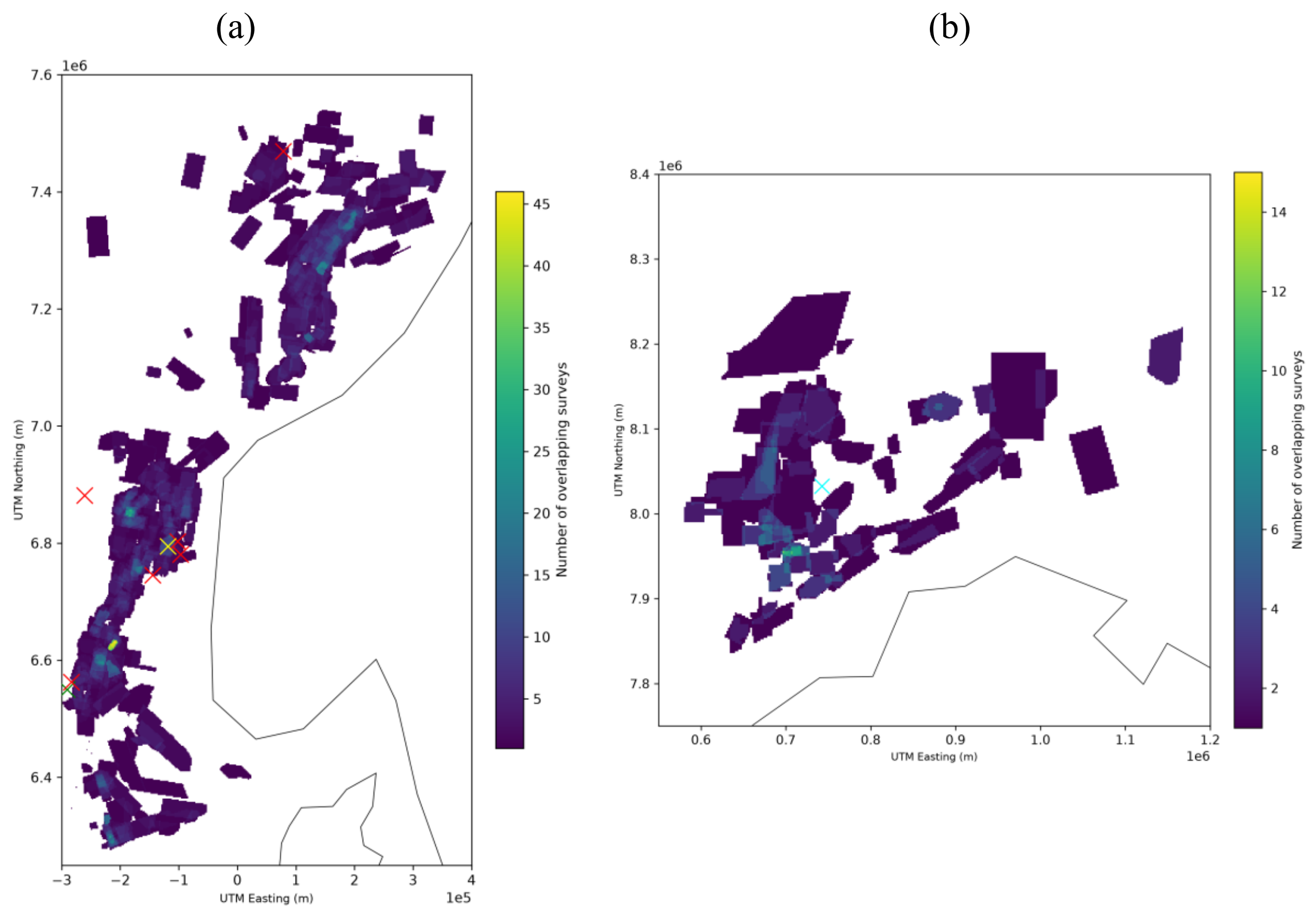

Figure 1: (a) Spatial distribution of the selected seismic cubes for the NCS-model for the North Sea/ Norwegian Sea (a) and the Barents Sea (b). Colors indicate the number of overlapping surveys per grid cell, highlighting regions of high acquisition density across the NCS. The crosses show the location of four of the test-datasets used in this study; salt segmentation in the Barents Sea (cyan); injectite mapping (green); flatspot mapping (yellow); package segmentation (red) in the North Sea/Norwegian Sea.

To avoid overrepresenting the data-rich zones and ensure broader geological and geographical coverage during pretraining, we sample training data more often from cubes in the low-density areas (Figure 1) than from cubes in the high-density areas.

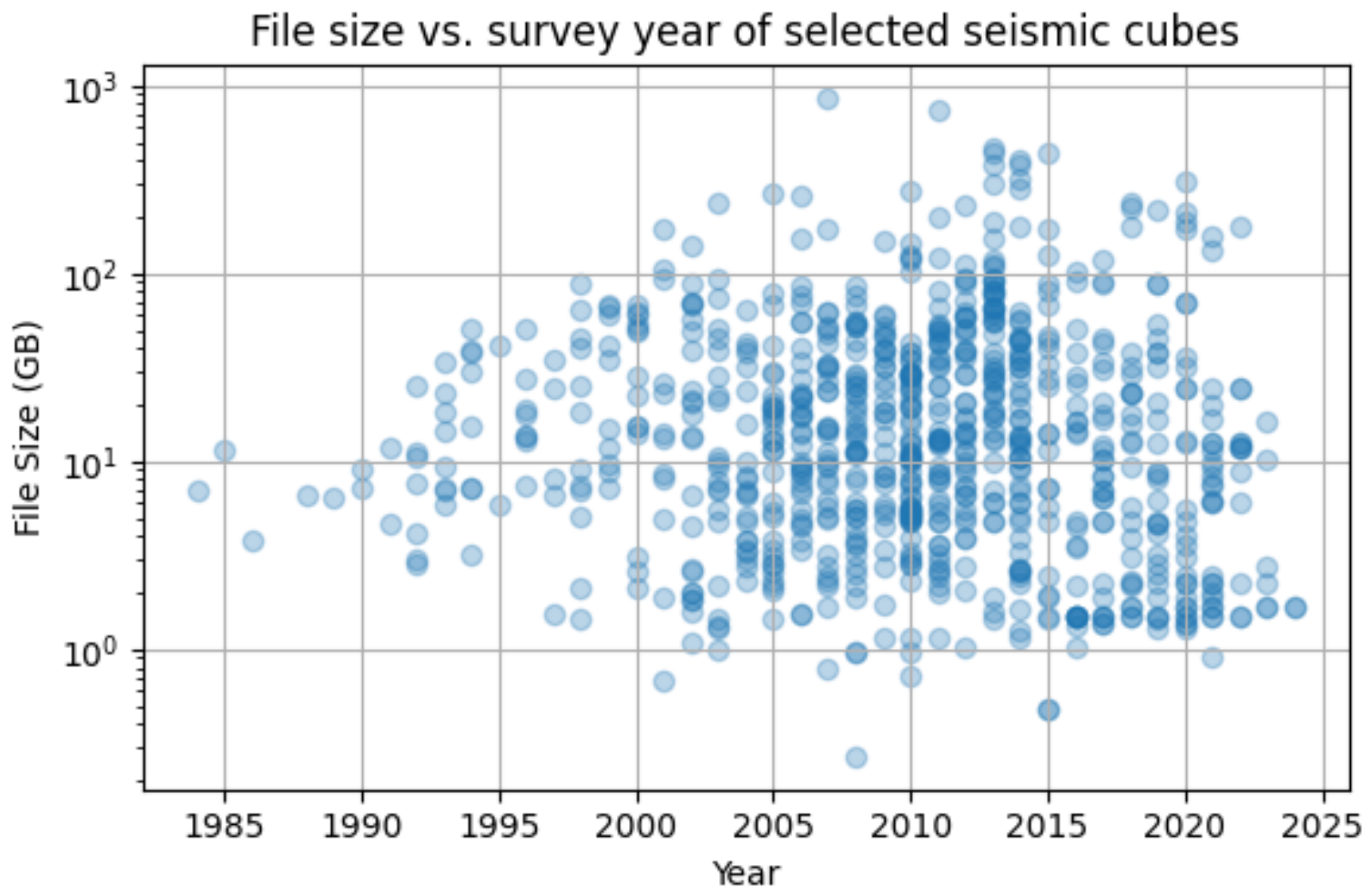

Figure 2: Distribution of seismic cube sizes (GB) as a function of acquisition year for the selected surveys.

To enable efficient random access to data during training, we convert all selected SEG-Y cubes to the *seismic-zfp* format (Wade, 2020). sThe seismic-zfp python library serves as a drop-in replacement for *segyio*, preserves header metadata, and enables low-overhead access to arbitrary sub-cubes. It uses ZFP compression, a highly efficient and near lossless compression. For efficient reading of data, we need to store the data on local storage on the compute nodes and therefore use a ZFP compression factor of 8 to fit more data onto the nodes. Training data are distributed evenly across high-performance computing (HPC) nodes to meet storage constraints. Each node stores and processes separate parts of the training data. Since the sizes of the 829 cubes are highly heterogenous (Figure 2), we split each cube into non-overlapping 1024 × 1024 × $N_z$ cubes (where $N_z$ is the number of samples in the depth/time dimension). The resulting 4601 sub-cubes are distributed evenly across the compute nodes, making it possible to store all the training data in the system memory across 4 nodes. Before training, each seismic cube is standardized to have unit variance, and amplitude values exceeding three standard deviations are clipped.

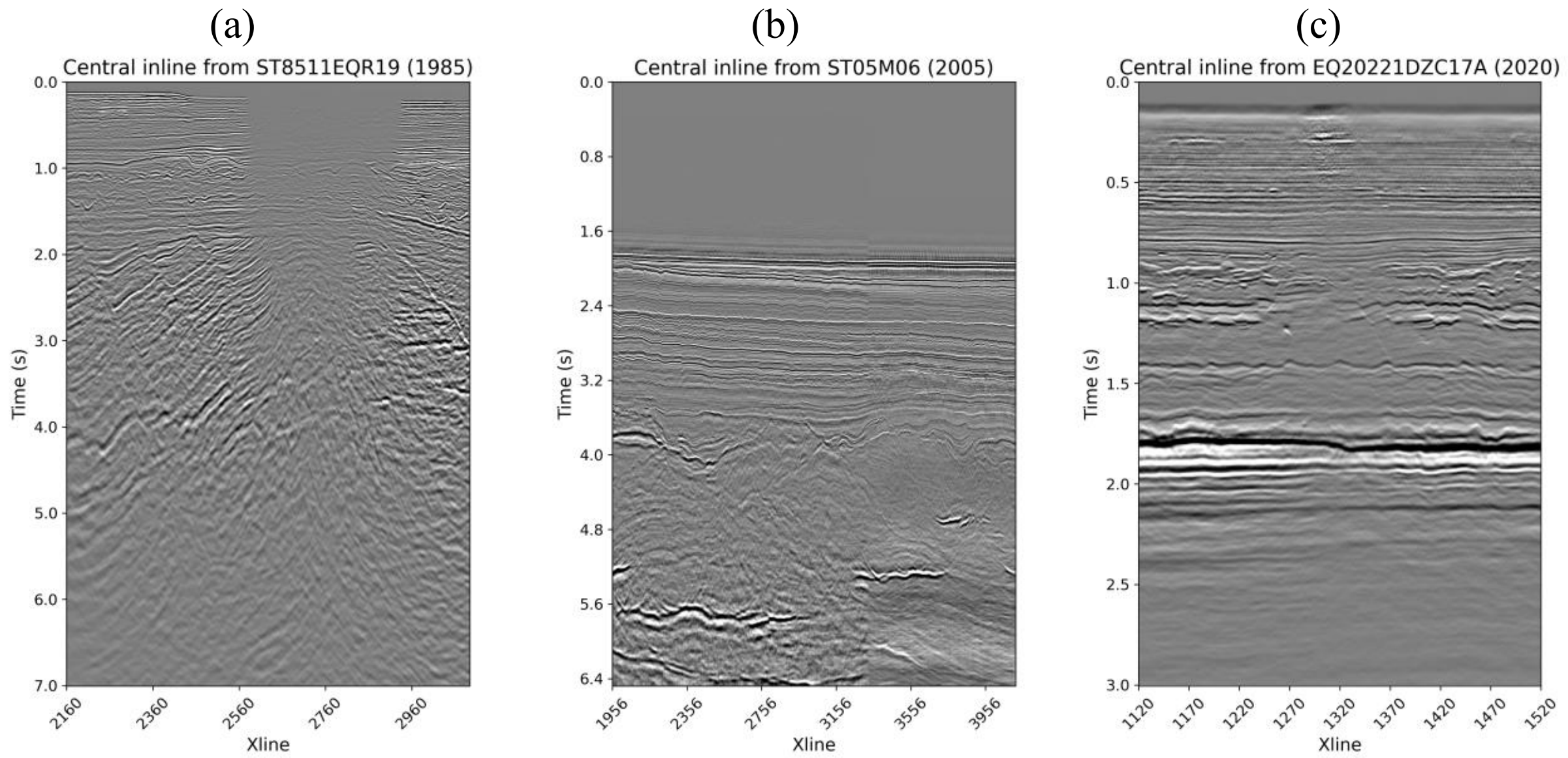

Figure 3: Representative inline sections illustrating the range of data quality present in the DISKOS repository. Examples include severely degraded data exhibiting acquisition footprint and low signal-to-noise ratio (a), moderate-quality legacy data with limited bandwidth and noise (b), and high-quality broadband data with strong reflector coherency (c).

## MODEL ARCHITECTURE AND TRAINING

We adopt the masked autoencoder (MAE) framework implemented with a ViT backbone as proposed by (He et al., 2022). Building on earlier work by (Ordoñez et al., 2024), we evaluate how alternative tokenization strategies, 2D, 2.5D, and 3D, affect the learned seismic representations when pretraining at NCS scale.

Each model variant is trained using the standard MAE reconstruction objective. As illustrated in Figure 4, inputs are extracted as 224 × 224 windows, with a 16 × 16 patch size for the 2D and 2.5D variants. For a 224 × 224 crop, the 2D input has 196 tokens. In the 2.5D formulation, we follow (Waldeland et al., 2025) by jointly providing the encoder with four views of the same local seismic neighborhood: an inline slice, a crossline slice, and two diagonal slices, which would correspond to 45° and 135° if the inline/crossline spacing is equal. We concatenate the four views giving us 784 tokens, which are processed jointly by the transformer.

For the 3D variant, we use 16 × 16 × 16 samples per patch extracted directly from volumetric blocks. However, a dense 224 × 224 × 224 block would contain 14 × 14 × 14 = 2744 volumetric tokens with 16-sample patch edges. To limit memory usage, we use a pillar-based sub-sampling strategy. The sub-volume is split into a 14 × 14 grid of non-overlapping vertical pillars. Each pillar is 16 × 16 samples laterally and spans the full 224-sample vertical extent, so it holds 14 stacked patches. For each training sample, we randomly select 40% of the pillars, which gives 78 × 14 = 1092 tokens, a sequence length comparable to the 2.5D variant. Every selected pillar keeps its complete vertical context, while the lateral context is sparse but spread over the whole 224 × 224 footprint. The selected pillars are concatenated into a single sequence, retaining each token's original coordinates in the sub-volume for the positional encoding. A new random subset of pillars is drawn for every sample, so different lateral locations are sampled throughout training.

All models use an 85% masking ratio, sampled uniformly at random over the flattened token sequence. In the 2.5D case, this masking is applied after the concatenation of patches across views, without enforcing a per-view masking quota, so the number of masked patches can vary across views within a sample. For the 3D variant, masking is applied after pillar sub-sampling.

Positional embeddings are adapted to the dimensionality and view composition of each variant. The 2D model uses standard 2-D sine-cosine positional encodings. The 2.5D model incorporates (x, y) sine-cosine embeddings combined with a cyclic orientation code that differentiates inline, crossline, and diagonal views, following the multi-view formulation of (Lee et al., 2023). For the 3D model, the encoder uses Lie Rotational Positional Encodings (LieRE) (Ostmeier et al., 2025), applying coordinate-dependent rotations to the queries and keys in each attention layer. Token coordinates refer to their original locations in the sub-volume, preserving the spatial arrangement of the sampled pillars. The 3D decoder uses fixed absolute 3D sine-cosine positional embeddings. All models share the same ViT-Base encoder (12 layers, hidden dimension = 768) and a lightweight 8-layer MAE decoder.

Pretraining is performed using 16 NVIDIA GH200 GPUs with BF16 mixed precision, flash-attention kernels, and a global batch size of 2048 (reduced to 1024 for the 3D model). We use a cosine learning-rate schedule with a 0.05 warmup ratio and a base learning rate of $1.5 \times 10^{-4}$, where the effective learning rate is the base learning rate times the total batch size divided by 256, similar to MAE pretraining (He et al., 2022). The pretraining budget is kept approximately constant across the three NCS variants, where each model is trained for 100 epochs, corresponding to approximately 1 million samples per epoch, totaling to around 250 GPU-hours per training run, measured as elapsed training time multiplied by the number of GPUs used.

Training progress is monitored using the MAE reconstruction loss, evaluated both on the training samples and on a held-out cube; in both cases, the loss decreases steadily over the course of pretraining for the three models.

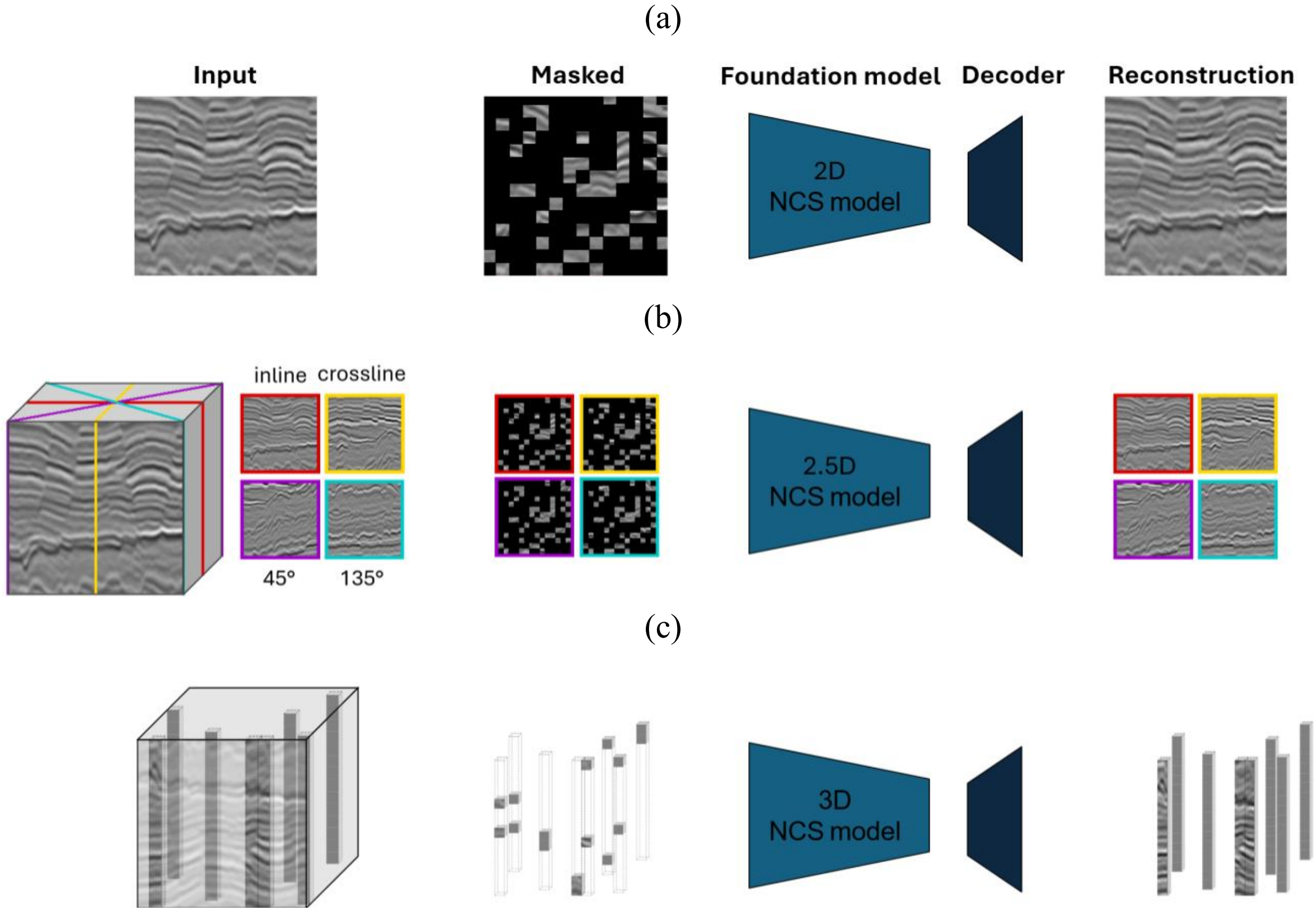

Figure 4: Overview of the input tokenization and reconstruction pipeline for the three NCS-model variants operating on crops of 224 vertical and horizontal extents. The 2D model (a) operates on a single 2D slice. The 2.5D model (b) jointly processes four views from the same local neighborhood: inline, crossline, and two diagonal slices (45° and 135° in the case of equal inline and crossline sampling). The 3D model (c) uses pillars sampled from volumetric blocks.

All models are initialized from ViT weights from MAE pretraining on ImageNet (Deng et al., 2009). For the 2D model no adaptation is required: it retains the original three-channel patch embedding, which is fed the seismic channel repeated three times, and its 2D positional encodings are left unchanged. For the 2.5D model, we average the input projection channels weights to map RGB to a single-channel seismic input and drop the original positional encodings. For the 3D model, we also average projection channel weights but expand and interpolate the 2D convolutional kernels to initialize $16 \times 16 \times 16$ volumetric patch embeddings; the original 2D positional encodings are removed. In our experiments with ViT-Base models, random initialization occasionally led to training collapse, whereas ImageNet initialization provided more stable optimization over the number of training epochs considered and allowed higher learning rates.

Training samples are drawn as random sub-cubes from the preprocessed seismic-zfp archive described above, enabling efficient I/O during large-scale pretraining.

## EVALUATION METHODOLOGY

We evaluate frozen pretrained encoders using two downstream classifiers: k-nearest neighbors (kNN) and a supervised linear probe. kNN tests local neighborhood structure in the embedding space; linear probing is included to assess whether the model ranking observed with kNN is preserved when using a trained classification head.

For kNN, we use $k = 5$ with Euclidean distance and uniform (majority) voting. Before both kNN classification and linear probing, embeddings are standardized per channel (z-scored) using a mean and standard deviation computed over the whole feature cube. For models evaluated with inline-crossline late fusion, this is done separately for each direction before the direction embeddings are concatenated.

The linear probe consists of a single linear layer trained using cross-entropy loss and full-batch Adam optimization for 1000 epochs (learning rate = $10^{-3}$, weight decay = $10^{-4}$), with predictions obtained by taking the argmax of the output logits. Neither protocol includes end-to-end encoder fine-tuning or a spatial decoder. This evaluates representation utility under limited supervision, without establishing the maximum downstream performance attainable by each model.

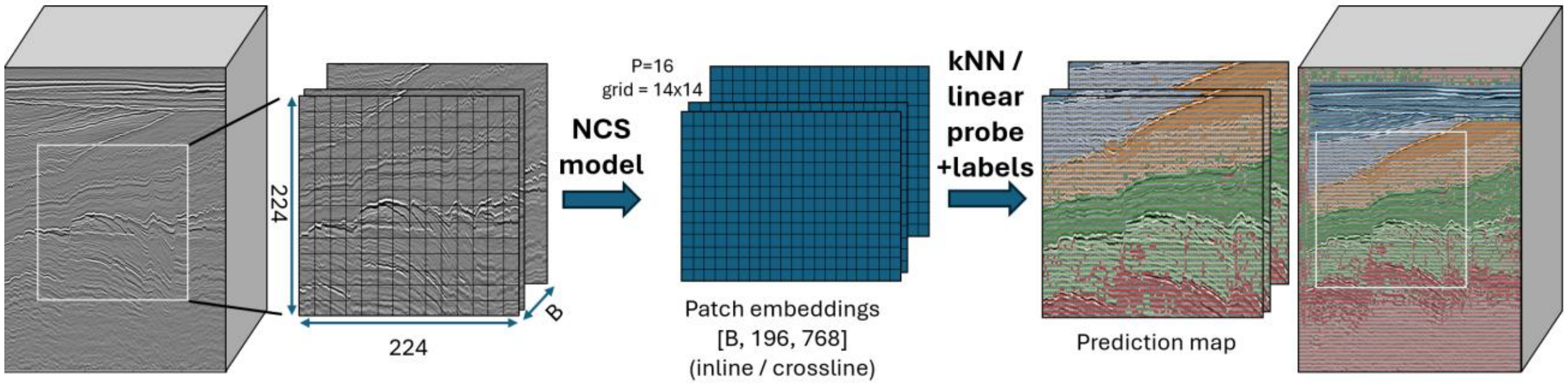

Figure 5: Patch-level feature extraction and classification using kNN or linear probing (2D example, package segmentation). A batch of B seismic crops of size 224 × 224 is partitioned into non-overlapping patches of size 16 × 16, yielding a 14 × 14 grid of 196 patch tokens per crop. The encoder produces patch embeddings of shape [B, 196, 768] for each orientation. Inline and crossline embeddings are concatenated at corresponding spatial locations, yielding 1,536-dimensional features. A kNN classifier or linear probe is fitted using a labeled support set, with each support embedding assigned the reference label at its patch-center coordinate by nearest-neighbor lookup on the label grid. The classifier assigns a class to each resulting feature vector. For visualization, each predicted label is assigned to the corresponding 16 × 16 patch region and mapped back to the seismic section.

Figure 5 illustrates the patch-level evaluation protocol. The encoder outputs a 768-dimensional embedding for each token, with one token per patch: 16 × 16 samples for the 2D and 2.5D models and 16 × 16 × 16 samples for the 3D model. These dimensions describe the patch grid rather than the full spatial context informing each embedding. For an input crop, this yields a grid of N token embeddings (N = 14 × 14 = 196 in the 2D case), which are used as the representation for all downstream tasks. Each embedding is associated with one reference label sampled at its patch-center coordinate using nearest-neighbor lookup on the label grid; reference labels are not aggregated over the patch. The kNN classifier and the linear probe are fitted on embeddings at the selected labeled support locations and then applied to embeddings extracted across the seismic volume.

For the 2D variant, we incorporate local 3D context via late fusion: inline and crossline embeddings are computed separately by the NCS encoder and concatenated at each patch location before inference. For the 2.5D variant, the encoder jointly processes four views, and the resulting patch embeddings are classified directly with kNN or linear probing. For the 3D variant, the encoder operates on volumetric blocks, and its 3D patch embeddings are likewise directly classified.

We evaluate representations on five interpretation benchmarks: four from the NCS and one out-of-basin benchmark, Penobscot. The benchmarks target distinct geological structures or stratigraphic phenomena and differ in their reference annotations, as described below.

1. **Salt segmentation (binary)**: Distinguishing salt bodies from background reflectivity. Reference labels were created by interpreting ten lines spread out across the seismic cube.
2. **Package segmentation (multi-class)**: Assigning stratigraphic units or seismic facies based on reflector character. Reference labels were derived from a set of complete key horizon interpretations, followed by conversion into package labels.
3. **Injectite segmentation (binary)**: Identifying sand injectites structures as visible on full stack data. The reference labels were obtained on five lines spread across the seismic cube and were model-derived proxy labels generated using a dedicated injectite detection model described in (Nguyen et al., 2024).
4. **Flatspot mapping (binary)**: Detecting amplitude anomalies consistent with fluid contacts. Reference labels were obtained from an interpretation of the reservoir outline of the seismic cube, restricted to the single depth slice at which the interpreted contact was most fully annotated.
5. **Penobscot facies classification (multi-class, out-of-basin)**: Assigning each sample to one of eight stratigraphic facies intervals spanning the full depth range of the volume. Reference labels were taken from the publicly available Penobscot interpretation (Baroni et al., 2020), in which a set of key horizons interpreted across the entire survey divides the volume into the eight intervals.

The package-segmentation benchmark comprises seven seismic cubes, each with a different number of facies classes, whereas the remaining tasks each use a single cube. Four of the benchmark data originate from publicly available NCS surveys, as indicated in Figure 1, and may overlap with the pretraining corpus in geographic location, although labels were never used during foundation model training.

Supervision is intentionally sparse: for Salt, Package, Flatspot, and Penobscot, we use 100 annotated support locations per class. We repeat downstream evaluation using five random seeds, each determining only the sampled support locations for those benchmarks. Locations are sampled uniformly without replacement. Each of the seven Package cubes is sampled independently, and the resulting scores are averaged across cubes. For Flatspot, sampling is restricted to the depth plane containing the largest number of annotated flatspot voxels. This plane is selected from the reference annotations and is identical across models and seeds. Within each seed, all models and both classification protocols use exactly the same support locations.

For Injectite, which exhibits strong class imbalance, the support set comprises all annotated locations on the middle inline of the evaluation grid. This support set is fixed across runs, so Injectite results have zero variability across seeds.

No other component of downstream evaluation is randomized: kNN is deterministic, and the linear probe uses fixed initialization and full-batch training. Consequently, variability across seeds reflects only the selection of labeled support locations.

| **Benchmark** | **Labels used for training** | **Scored locations** | **Class proportions** |
|---|---|---|---|
| Salt | 200 (100 × 2 classes) | 27,347 | C0 78.2%; C1 21.8% |
| Package | 600–900 per cube<br>(100 × 6–9 classes) | 1,692,337<br>(7 cubes) | Not reported |
| Injectite | 11,623 (C0 11,513; C1 110) | 58,230 | C0 98.9%; C1 1.1% |
| Flatspot | 200 (100 × 2 classes) | 9,546 | C0 57.6%; C1 42.4% |
| Penobscot | 800 (100 × 8 classes) | 43,848 | Not reported |

Table 1: Training labels, scored locations, and class composition for the five benchmarks. Class proportions are only reported for binary benchmarks at the scored locations, where C1 denotes the class of target structure and C0 the background. These proportions do not describe the sparsely labeled training set.

Scores are computed at all annotated locations on the evaluation grid, sampled at intervals of 16 samples along each axis, including the locations used for support. Unannotated locations are excluded. Flatspot scoring is restricted to the selected depth plane, while Injectite scoring includes the annotated inline used for training. Table 1 summarizes class composition at the scored locations, which is distinct from the composition of the labeled support set. They are omitted for the multi-class benchmarks because Package comprises seven cubes, each with its own set of six to nine classes and Penobscot has eight classes, so a per-class breakdown would not fit concisely in the table.

Performance is reported as mean Intersection-over-Union (mIoU), calculated for each evaluation seed by averaging the class-specific intersection-over-union scores within each benchmark cube. We additionally report macro precision and recall characterizing false-positive and false-negative class assignments, and HD95, expressed in evaluation-grid units, to measure discrepancies between predicted and reference boundaries, with lower values indicating better boundary agreement. HD95 is computed for the target class in binary benchmarks and averaged across classes in multiclass benchmarks. For Package, each metric is averaged equally across the seven cubes. Each benchmark result is reported as the mean and standard deviation across five evaluation seeds.

## BENCHMARK EVALUATION RESULTS

Figure 6 compares the mIoU achieved by the proposed NCS-model variants against three baselines: (i) a ViT-MAE pretrained on ImageNet, (ii) DINOv3 (Siméoni et al., 2025), a state-of-the-art frontier vision foundation model pretrained on large-scale natural image data, and (iii) the SFM-B model from (Sheng et al., 2025) pretrained on a globally aggregated seismic dataset. All three baselines follow the same late-fusion strategy as NCS-2D: embeddings are extracted separately from inline and crossline slices and concatenated at each evaluated location before kNN classification or linear probing.

For all models, including ours, amplitudes in seismic cubes are normalized to unit variance and the ones exceeding three standard deviations are clipped. For the natural-image baselines, the single-channel seismic input is replicated across three channels, and amplitudes are rescaled to lie within [0, 1] before normalization using the ImageNet channel-wise means and standard deviations.

The ImageNet-pretrained ViT-MAE baseline has the lowest average performance of all models under both kNN and linear probing, with the largest gaps on Package segmentation and Penobscot under kNN. DINOv3 is a much stronger starting point and achieves the best Salt performance under both protocols, although the differences between models on this task are within seed variability. On average, however, DINOv3 remains behind most seismic-pretrained models: SFM-B, NCS-2D and NCS-2.5D achieve higher averages under both protocols, while DINOv3 and NCS-3D perform comparably.

With kNN, NCS-2.5D achieves the highest average mIoU followed by NCS-2D and SFM-B. Linear probing raises the five-benchmark average for every model with the highest average remaining NCS-2.5D followed by SFM-B and NCS-2D. These rankings are based on means over five evaluation seeds. The generally small standard deviations, particularly under linear probing, indicate limited seed-to-seed variability, although kNN is less stable for some model-task combinations, especially on Flatspot.

The advantage of NCS-2.5D is not restricted to the in-basin benchmarks, as it also achieves the highest performance on the out-of-basin Penobscot dataset. NCS-3D shows a different behavior. It obtains lower overall averages than the 2D and 2.5D variants, with weaker Package and Flatspot performance under kNN and a pronounced weakness on Injectite under linear probing. This indicates that the way we have learned the 3D representations may not be optimal to provide uniform benefits across the considered interpretation tasks. Nevertheless, NCS-3D transfers particularly well to the out-of-basin Penobscot and the Flatspot benchmarks under linear probing.

Supplementary Table S1 provides the numerical mIoU results shown in Figure 6, while Tables S2-S4 report macro precision, macro recall, and HD95. NCS-2.5D achieves the highest average precision and recall under both kNN and linear probing, consistent with its leading average mIoU. Boundary agreement is more task-dependent: NCS-2.5D achieves the lowest HD95 on Package, Flatspot, and Penobscot under both protocols, but one of the highest HD95 values on Injectite with kNN, where it has the highest mIoU. The latter illustrates that strong segmentation overlap can coexist with larger boundary discrepancies.

Overall, these results suggest that within the NCS family, the 2.5D multi-view formulation captures useful volumetric context and offers an effective approach to large-scale seismic pretraining.

(a)

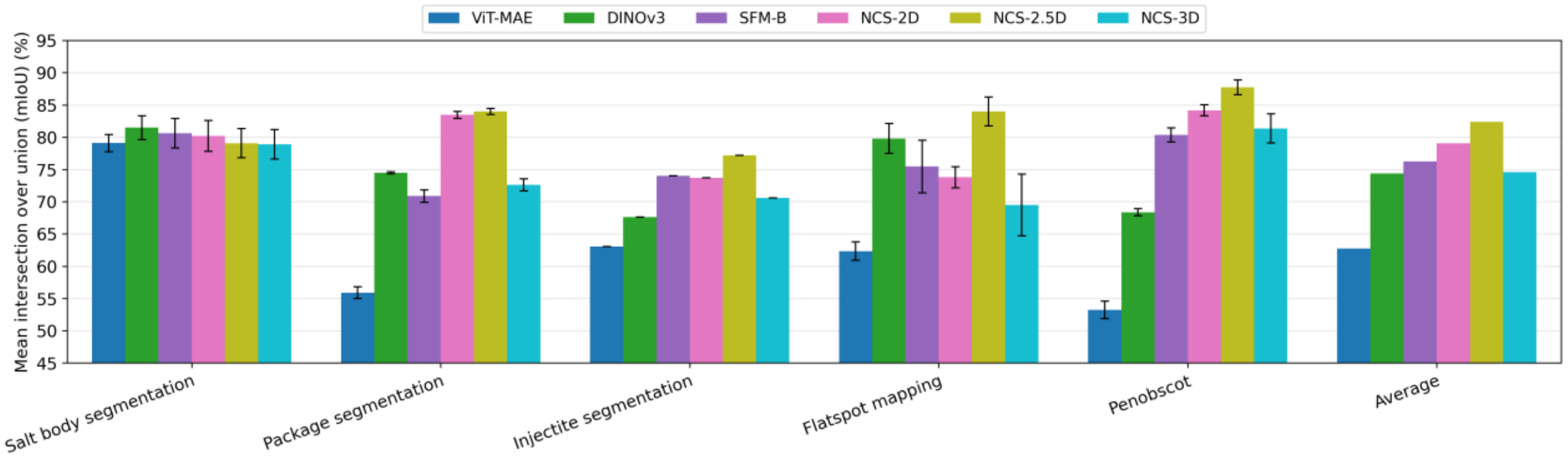

(b)

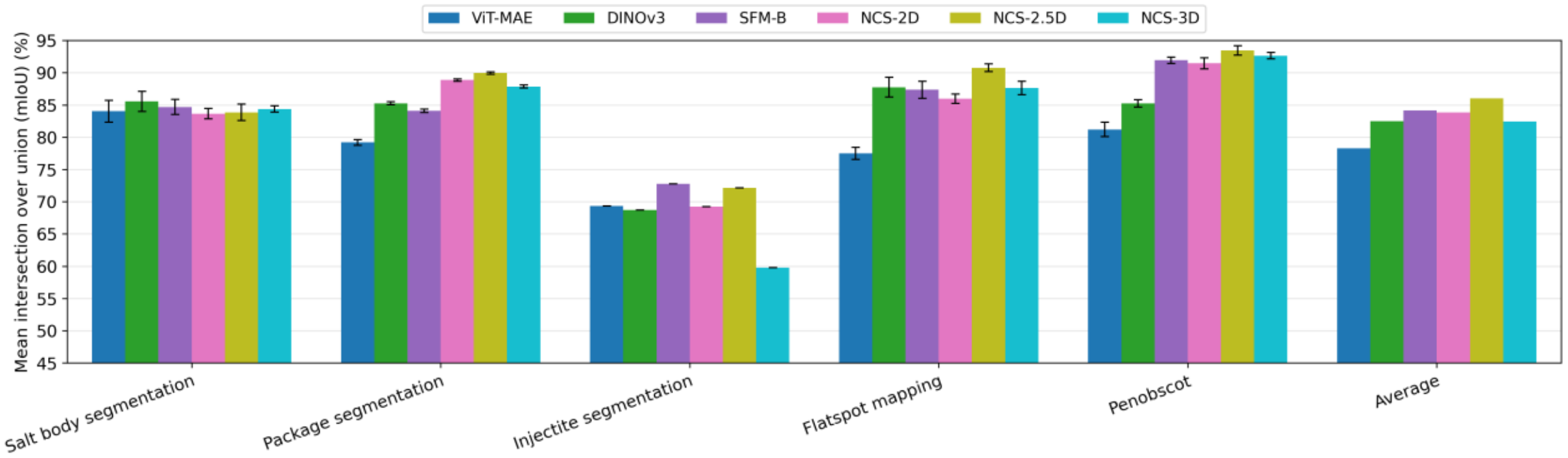

Figure 6: Benchmark evaluation of pretrained models across five interpretation tasks for kNN (a) and linear probing (b). Models include a ViT-MAE pretrained on ImageNet, DINOv3, the globally trained seismic foundation model (SFM-B) from (Sheng et al., 2025), and the proposed NCS-models. Bars show mean mIoU over five evaluation seeds; black error bars indicate standard deviations across these seeds. The final group shows the unweighted average of the five benchmark means, without error bars.

Figure 7 complements the quantitative evaluation by illustrating differences in the spatial coherence and geological consistency of predictions produced by the evaluated models using kNN with sparse supervision.

On salt segmentation, all models recover the main body consistent with the small mIoU spread in this task, but DINOv3 exhibits cleaner, more laterally continuous boundaries with fewer isolated misclassified patches in the surrounding sediments. For package segmentation, the NCS models better preserve stratigraphic continuity and produce sharper, horizon aligned transitions between units, introducing less class leakage across reflectors compared to the ViT-MAE and SFM baselines, especially for the shallower units. For injectite mapping, seismic pretrained models recover a larger fraction of true positives within the labeled interval and produce more spatially coherent detections along the thin, laterally persistent injectite signal, with the strongest continuity observed for the NCS models. This behavior suggests improved sensitivity to the subtle amplitude and textural cues that characterize injectites. For flatspot mapping, the DINOv3 and the 2.5D NCS models localize the flatspot anomaly more consistently and concentrate the activation within the reservoir outline.

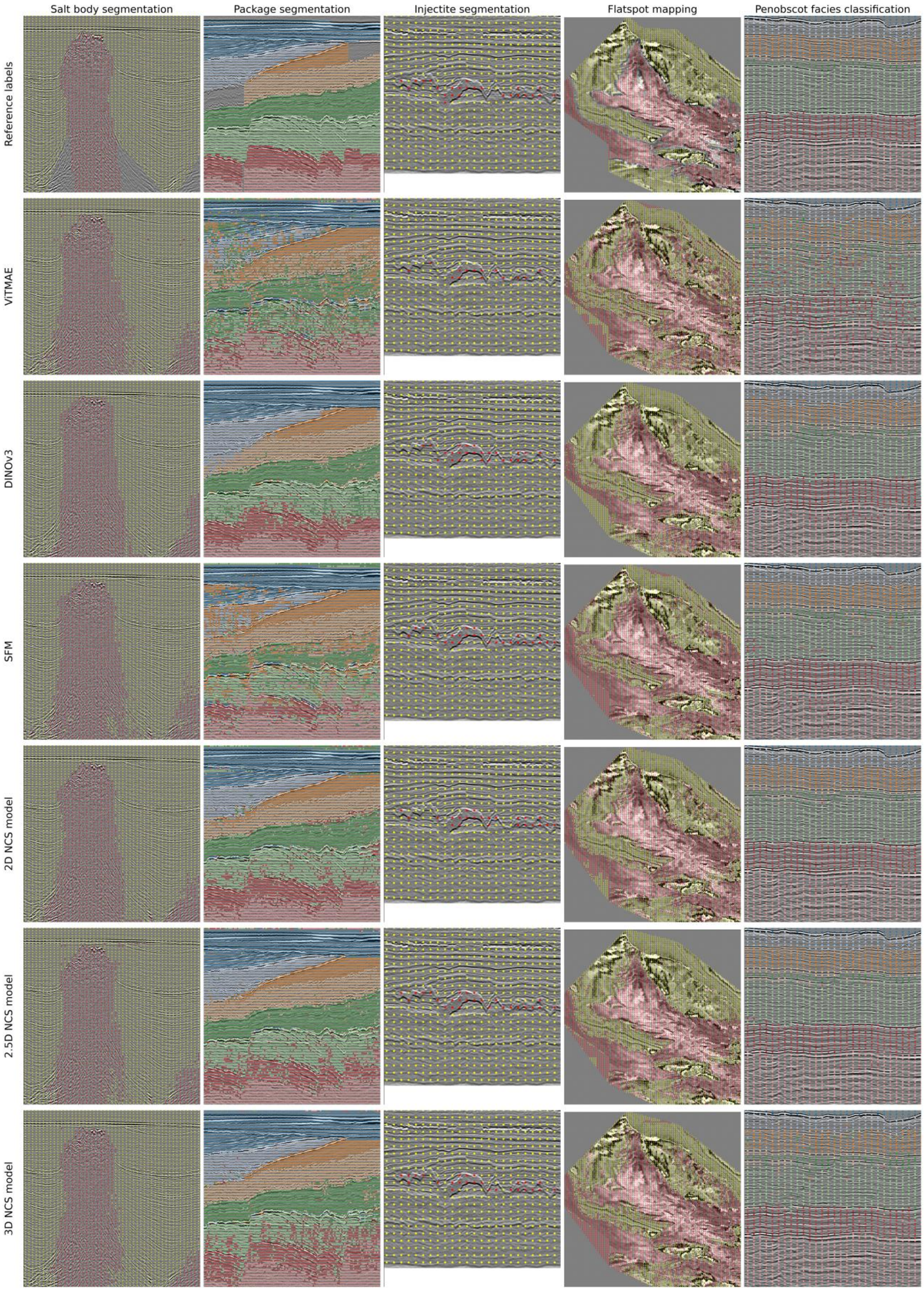
Salt body segmentation
Package segmentation
Injectite segmentation
Flatspot mapping
Penobscot facies classification
Reference labels
ViTMAE
DINOv3
SFM
2D NCS model
2.5D NCS model
3D NCS model

Figure 7: Qualitative results across five downstream interpretation tasks obtained with kNN. Columns show salt body segmentation, stratigraphic package segmentation, injectite segmentation, flatspot mapping and Penobscot facies classification. Rows compare reference labels with predictions from the evaluated baselines (ViT-MAE, DINOv3, SFM) and the proposed NCS models (2D, 2.5D, and 3D). For the binary tasks (salt and injectite), red indicates the presence of the target structure and yellow its absence. For package segmentation and Penobscot, colors denote distinct stratigraphic units. For flatspot mapping, the panel shows a depth slice where red highlights the detected flatspot.

## EFFICIENCY AND MODEL CHARACTERISTICS

We report compute and memory proxies for the NCS-2D, NCS-2.5D, and NCS-3D variants (Table 2).

| **Metric** | **NCS-2D** | **NCS-2.5D** | **NCS-3D** |
|---|---|---|---|
| Model parameters (M) | 85.80 | 85.86 | 88.21 |
| Peak GPU mem FP32 (MB) | 345.13 | 367.89 | 766.89 |
| Peak GPU mem FP16/BF16 (MB) | 184.41 | 195.87 | 578.13 |
| Compute (GMac) | 16.87 | 66.91 | 242.19 |
| Inference time (s) | 19.0 | 19.6 | 12.4 |

Table 2: Characteristics of the NCS model variants. Parameter counts are reported in millions (M). Compute (GMac) and peak GPU memory (MB), including model parameters are measured for a single forward pass with patch size 16 and batch size 1, using input sizes of 224 × 224 (2D), 4 × (224 × 224) views fused into one sequence (2.5D), and 224 × 224 × 224 (3D). Inference time is end-to-end runtime for embedding extraction on a 1024 × 1024 × 1024 seismic cube on a single RTX 4000 (20 GB), using FP16 for all models.

While parameter counts are comparable across models, the 2.5D variant incurs higher computational cost and peak GPU memory (including model parameters) than 2D but remains substantially less demanding than the evaluated 3D configuration. We additionally report end-to-end inference time for embedding extraction on a full 1024 × 1024 × 1024 cube on a single GPU using FP16. All variants produce features at the same output sampling interval, and the NCS-2D timing includes both inline and crossline feature extraction and their concatenation. Batch size and dataloader parallelism are adjusted per variant to respect memory limits; these runtimes reflect practical throughput under feasible settings.

## EXAMPLE: USING THE NCS-MODEL FOR INTERACTIVE INTERPRETATION

To illustrate how the learned representations support practical seismic interpretation, we apply the similarity-search approach proposed in (Waldeland et al., 2025), which enables user-guided mapping of geological structures. For each of the examples shown in Figure 8, the NCS-2.5D encoder was used to precompute patch embeddings within the seismic cube at a sampling interval of four seismic samples and the similarity search was operated on the resulting feature volume. The user provides positive example clicks corresponding to the structure of interest, and their embeddings are averaged to form a query vector. When negative examples are provided, their mean embedding is subtracted from the mean positive embedding. A similarity score is then computed at each location as the dot product between its embedding and this query vector, producing a volumetric map that highlights patterns similar to the positive examples and dissimilar to the negative examples (Waldeland et al., 2025). The similarity computation operates directly in the embedding space and is accelerated with SIMD-optimized similarity metrics and quantization through the vector search library Usearch (Vardanian, 2023), leading to full-volume similarity maps generated in seconds.

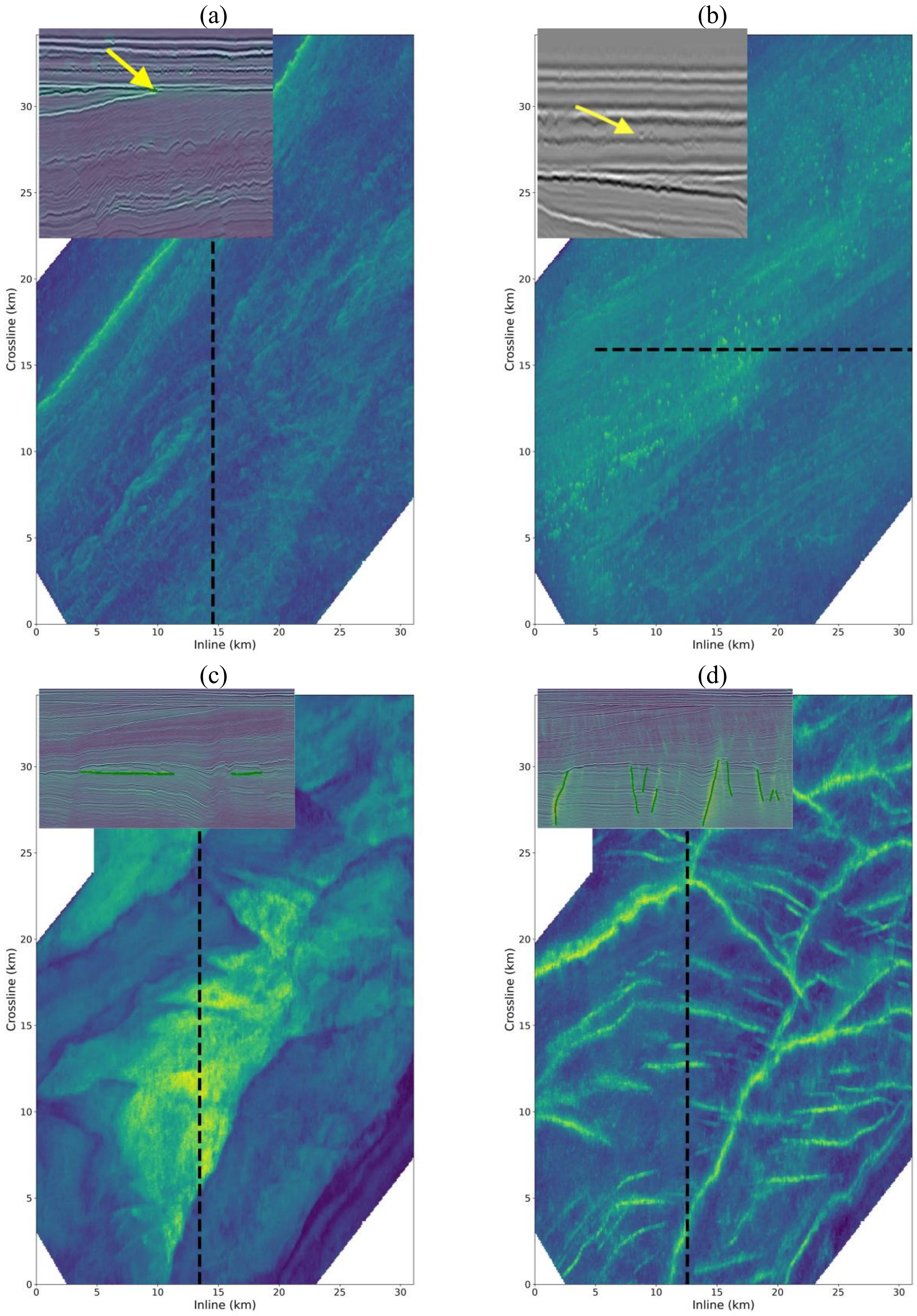
(a)
(b)
(c)
(d)
Crossline (km)
Inline (km)
0
5
10
15
20
25
30

Figure 8: Similarity maps produced by the 2.5D NCS-model for four representative geological features: (a) truncation point, (b) boulders in the near seabed sediments, (c) flatspot, and (d) faults in the reservoir zone. Yellow highlights regions whose embeddings closely match the user-selected positive examples; blue indicates low similarity. The black dashed line indicates the seismic section from which the examples were selected. Insets (top) show the corresponding seismic sections: in (a, b) the single positive example click used for similarity search is indicated by an arrow, whereas in (c, d) the positive examples, lines of points, are highlighted in green.

The resulting similarity maps shown in Figure 8a and Figure 8b use one positive exemplar; the other panels use the positive selections shown in green. No negative exemplars were used in any of the four examples. Truncation points (Figure 8a) are highlighted with strong, localized contrast along reflector terminations, and boulders in the near seabed sediments (Figure 8b) are retrieved as clustered, high similarity anomalies consistent with their seismic expression. The flatspot example (Figure 8c) yields a laterally continuous response mostly confined to the target level, while faults (Figure 8d) are recovered as connected networks with good continuity. These examples illustrate how the feature space learned by NCS-2.5D can support rapid, interactive interpretation workflows, enabling geoscientists to explore large 3D cubes with minimal manual labelling.

## DISCUSSION

Recent work has begun to consolidate the concept of training seismic foundation models, with efforts centering on masked autoencoding using ViT backbones (Sheng et al., 2025; Pham et al., 2025; Sansal et al., 2025). These studies show that pretraining on seismic cubes can improve performance across a range of downstream interpretation tasks. Building on this context, we return to the central methodological question posed in the Introduction: how do foundation models pretrained on the NCS archive perform relative to available natural-image and seismic baselines? The discussion below interprets our results in light of this question and, in doing so, directly addresses the three aims of this study: establishing region-specific foundation models trained at continental scale on real industrial data, comparing practical 2D, 2.5D, and 3D configurations under similar pretraining budgets, and enabling follow-up research through the release of pretrained models. We then discuss how these outcomes depend on methodological choices in representation design and evaluation.

**Main findings**

The benchmark results provide an empirical answer to the central question. Models pretrained on natural images, including both a ViT-MAE pretrained on ImageNet (He et al., 2022) and a state-of-the art frontier model such as DINOv3 (Siméoni et al., 2025) trained on large-scale natural image collections, show task-dependent transfer to seismic interpretation. The ImageNet-pretrained ViT-MAE achieves lower average mIoU than DINOv3 and all seismic-pretrained models under both evaluation protocols. DINOv3 achieves the highest mIoU on Salt and performs strongly on Flatspot but performs less well than other seismic models (SFM-B and NCS-2.5D) on Injectite and Penobscot. This pattern suggests that the domain gap between natural images and migrated seismic remains relevant to downstream performance, although its effects vary across interpretation tasks. Seismic-pretrained models achieve stronger results on average, with SFM-B (Sheng et al., 2025) outperforming the natural-image baselines in average mIoU and NCS-2.5D achieving the highest average under both kNN and linear probing. Taken together, these findings support the core premise behind seismic foundation models: although generic vision pretraining can provide useful representations, pretraining directly on seismic data can improve their effectiveness across a range of interpretation tasks.

A second important outcome is that NCS-2.5D achieves measurable gains over the globally pretrained seismic baseline. It achieves the highest average mIoU across the benchmark suite, with improvements over SFM-B on flatspot mapping and stratigraphic package segmentation under both evaluation protocols. These tasks demand sensitivity to subtle amplitude variations and complex reflector geometries, and the observed gains suggest that pretraining on the NCS archive helps capture patterns relevant to these interpretation tasks. At the same time, the competitiveness of the global baseline on salt segmentation and injectite mapping (linear probe) highlights the task-dependent nature of these advantages. The strong performance of NCS-2.5D on Penobscot further demonstrates that its usefulness extends beyond the NCS pretraining region. These results suggest that large-scale pretraining, even when restricted to the NCS, can yield useful representations for interpretation tasks outside the pretraining region. However, differences in architecture, pretraining data, and training recipe prevent attributing the gains over SFM-B specifically to the scale or regional focus of pretraining. Broader out-of-basin evaluation is needed to establish how this transfer varies across geological settings and acquisition and processing conditions.

**Representation design and practical implications**

The comparison of practical model configurations under similar pretraining budgets highlights accuracy and efficiency trade-offs in representation design. The 2.5D formulation, which fuses four views (inline, crossline, and two diagonal slices) into a single token sequence and jointly masks them, achieves the highest average mIoU under both evaluation protocols. This suggests that multi-view representations capture useful local 3D context, potentially because complementary orientations expose structural cues and amplitude patterns without requiring dense volumetric tokenization. From an engineering perspective, these results position 2.5D as a practical approach to large-scale seismic pretraining: it maintains a comparable model size while requiring less computation and peak GPU memory per evaluated input than the profiled 3D configuration (Table 2). These per-input costs should be distinguished from full-volume processing throughput, for which NCS-3D achieves the shortest embedding-extraction time under the tested settings.

The 3D model shows task-dependent performance and its average mIoU is lower than that of the 2D and 2.5D variants. These results should be interpreted in light of the practical training constraints in our study and the differences between our setup and that from (Sansal et al., 2025). In our case, pretraining uses an NCS-only corpus, and the three model variants are compared under similar pretraining budgets. When processing dense mini-cubes, memory constraints required a substantial reduction in batch size, reducing the number of samples contributing to each update. To keep training feasible, we therefore sampled random pillars rather than processing complete dense mini-cubes. Pillar sampling aligns with seismic-zfp chunking and preserves the full vertical extent at selected lateral locations while reducing the number of volumetric tokens processed. However, this subsampling may limit how fully the model exploits volumetric relationships compared with dense 3D pretraining. Our results therefore characterize a practical 3D configuration under the available resources, leaving open whether denser sampling or a larger training budget would improve its performance relative to the multi-view formulation.

Evaluating the different tokenization strategies also required selecting positional encoding schemes that are compatible with each input representation. In practice, this means encoding in-plane position, view identity, and volumetric coordinates differently across 2D, 2.5D, and 3D variants. Although these schemes were chosen to suit each representation, they may also contribute to the observed performance differences, preventing attribution of those differences to tokenization alone. An important direction for future ablations is therefore to disentangle the contributions of tokenization, input sampling, and positional encoding to the observed gap between the 2.5D and 3D configurations.

Beyond benchmark metrics, the interactive similarity-search examples provide qualitative evidence that the learned embedding space captures patterns relevant to seismic interpretation. Following Waldeland et al. (2025), user-provided exemplars enable retrieval of visually and geologically similar patterns and can support rapid delineation of structures in a human-in-the-loop setting. This use case complements the supervised benchmarks by illustrating how the representations can support interpreter-guided exploration without training a task-specific prediction head.

A practical contribution of this work is the release of pretrained model checkpoints and accompanying inference code, enabling researchers to use the learned representations without repeating large-scale pretraining. We do not redistribute the underlying DISKOS seismic volumes; access remains governed by DISKOS terms, including any associated data access and transfer costs. Repository-scale seismic training involves substantial data preparation and I/O engineering, which can limit its accessibility. By documenting the DISKOS-derived selection protocol and training procedure and making the pretrained models available, we aim to facilitate their reuse in seismic interpretation workflows and support further research on regional seismic foundation models.

**Limitations**

All reported benchmark scores follow a deliberately constrained evaluation protocol: we keep the pretrained backbone frozen, use kNN or a linear probe, and assess performance on four NCS interpretation benchmarks and one out-of-basin benchmark. Together, these choices prioritize comparability across models and focus the evaluation on the usefulness of frozen representations, limiting the influence of complex task heads and excluding label-driven adaptation of the encoder. Accordingly, our conclusions should be interpreted as evidence about representation quality under standardized evaluation protocols on representative, but necessarily incomplete, interpretation use cases. More expressive task heads and end-to-end fine-tuning could improve absolute mIoU, although the relative ordering of models may also change. Furthermore, interpretation labels and benchmark definitions involve uncertainty and subjectivity, including the use of model-derived proxy labels for Injectite, which can affect measured performance and obscure small differences between models.

All NCS models were initialized from ImageNet MAE weights, and no randomly initialized model was trained to convergence for comparison; the contribution of this initialization to downstream performance is therefore not quantified.

Spatial resolution imposes a further limitation on both the learned representations and their evaluation. Patch-based embeddings summarize information over a finite spatial support, which may make thin geological bodies, closely spaced reflectors, and subtle discontinuities difficult to distinguish, particularly when they occupy only a small fraction of a patch. The spatial granularity of the reference annotations also determines which details can be assessed reliably: structures or boundary variations that are not resolved in the labels cannot be adequately evaluated. Consequently, the reported benchmark scores do not establish reliable detection or precise delineation of fine-scale geological structures. Dedicated evaluation using detailed annotations and metrics sensitive to small structures and boundary accuracy would help characterize these limitations. Smaller patches or multiscale representations could also be investigated, while accounting for their additional computational cost.

Data locality also limits the interpretation of the results. Pretraining is restricted to the NCS archive, and the four NCS benchmarks may overlap geographically with the pretraining corpus. Although task-specific labels were not used during foundation model pretraining, performance on these benchmarks does not establish transfer to geographically unseen surveys. Penobscot provides evidence of transfer beyond the NCS, but a single external survey is insufficient to establish broad generalization across geological settings and acquisition and processing conditions. Establishing a shared, publicly accessible benchmark suite with documented annotations, predefined data splits, and geographically independent surveys across diverse basins and interpretation tasks would support more systematic assessment of transfer and comparison of seismic foundation models.

The NCS archive spans multiple acquisition vintages and processing generations, from legacy surveys to more recent acquisitions. Including both time-migrated and depth-migrated full-stack cubes further broadens this diversity but also introduces heterogeneity in geometry and imaging characteristics. Exposure to both domains may help the models learn representations useful across different imaging conditions, although these differences could also make learning shared representations more challenging. The present experiments do not isolate the effect of mixing the two domains. A controlled comparison of time-only, depth-only, and mixed pretraining corpora, using comparable data volumes and training budgets, would help assess this effect and inform the development of models intended for reuse across industrial imaging workflows.

The present study remains limited to migrated full-stack amplitude volumes. Extending pretraining and evaluation to other seismic inputs, such as angle stacks, or incorporating multimodal supervision from well data would broaden the scope of future investigations.

**Future directions**

A natural question is whether seismic foundation models will follow scaling trends analogous to those observed in natural-image pretraining. Building on earlier seismic foundation model studies (Sheng et al., 2025; Pham et al., 2025; Sansal et al., 2025), the NCS corpus provides a substantial resource for investigating this question, although survey overlap and multiple processing or reprocessing variants introduce redundancy. The strong performance of NCS-2.5D demonstrates the effectiveness of the tested configuration following large-scale seismic pretraining but does not isolate the contribution of corpus size or establish a scaling law with respect to data volume or model capacity. The current experiments also do not establish whether returns have saturated, leaving open whether additional data or larger models would yield further improvements. Establishing scaling relationships will require controlled studies that vary corpus size, redundancy and diversity filtering, and model capacity, while accounting for the practical constraints of 3D I/O and memory. In this setting, effective diversity and curation may be as important as raw terabytes.

Dataset curation and sampling are central determinants of what a seismic foundation model learns. The DISKOS repository is heterogeneous in survey footprint, vintage, processing, and imaging quality, and it is spatially imbalanced with strong redundancy in data-rich areas. Without accounting for survey overlap, the training distribution can overrepresent frequently imaged areas and repeated structural patterns, potentially favoring local acquisition and processing characteristics. The density-aware sampling strategy used here aims to mitigate this imbalance, but geographic coverage alone does not ensure diversity in seismic appearance or geological content. A natural extension is balanced sampling based on visual features, for example by maintaining a lightweight embedding model and prioritizing underrepresented regions of the embedding space during pretraining. A practical template is the clustering-based automatic curation pipeline of (Vo et al., 2024), which uses hierarchical k-means clustering of self-supervised embeddings followed by balanced sampling across clusters. Such a strategy could help reduce redundancy and increase exposure to less common patterns, potentially complementing metadata and spatial density as indicators of geological and imaging diversity.

Masked autoencoding is an effective and simple baseline for seismic pretraining, but its learning signal is anchored in reconstructing masked inputs in pixel space, which may encourage sensitivity to local texture and amplitude variations that are not always relevant to interpretation. In seismic data, acquisition and processing artifacts contribute to these variations alongside geological structures, and similar amplitude patterns can admit different geological explanations. Our results also show that DINOv3 provides stronger frozen representations than ImageNet-pretrained MAE, motivating exploration of alternative pretraining objectives, although differences in training data and training recipe prevent attributing this advantage to the objective alone. Self-distillation methods in the DINO family (Caron et al., 2021; Oquab et al., 2023; Siméoni et al., 2025), or joint-embedding learning with explicit regularization as in LeJEPA (Balestriero and LeCun, 2025), offer promising alternatives to pixel reconstruction. Such approaches could help learn stable structural and stratigraphic representations across acquisition vintages and processing variants. Their potential to improve robustness and downstream transfer should be assessed using a common seismic corpus, comparable backbones, and similar pretraining budgets.

Looking forward, the most direct next steps are to expand evaluation across basins and contribute to a shared benchmark suite, study the effects of pretraining scale and corpus diversity, improve curation through feature-based clustering and balanced sampling, and compare alternative pretraining objectives using the same seismic data and similar training budgets.

# CONCLUSION

This study shows that large-scale self-supervised pretraining on the NCS archive can yield effective representations for seismic interpretation, with benefits extending beyond the pretraining region. Although natural-image models remain competitive on some tasks and no model is uniformly best across tasks and metrics, NCS-2.5D achieves the highest average mIoU under both kNN and linear probing. Its leading performance on Penobscot provides evidence of transfer to a geographically separate survey. Within the NCS family, its advantage under similar pretraining budgets supports the multi-view formulation as an effective configuration for large-scale seismic pretraining. These findings position regionally pretrained models as strong candidates for seismic interpretation, while leaving the contributions of data scale, regional coverage, and model configuration to be disentangled. The released pretrained checkpoints and inference code provide reusable resources for applied interpretation and further comparative research.

# DATA AND MATERIAL AVAILABILITY

Pretrained NCS-model checkpoints and the inference pipeline are publicly available through: https://github.com/NorskRegnesentral/NCS_models. The underlying seismic volumes are not redistributed by the authors because access to the Norwegian Continental Shelf data is governed by the DISKOS data-sharing terms. These data are openly accessible through DISKOS, but any costs associated with data access and transfer remain the responsibility of the user.